# Perceiving the Social: A Multi-Agent System to Support Human Navigation in Foreign Communities

*Victor V. Kryssanov, Ritsumeikan University, Japan*

*Shizuka Kumokawa, Ritsumeikan University, Japan*

*Igor Goncharenko, 3D Incorporated, Japan*

*Hitoshi Ogawa, Ritsumeikan University, Japan*

**Abstract.** This paper describes a system developed to help people explore local communities by providing navigation services in social spaces created by the community members via communication and knowledge sharing. The proposed system utilizes data of a community's social network to reconstruct the social space, which is otherwise not physically perceptible but imaginary, experiential, yet learnable. The social space is modeled with an agent network, where each agent stands for a member of the community and has knowledge about expertise and personal characteristics of some other members. An agent can gather information, using its social "connections," to find community members most suitable to communicate to in a specific situation defined by the system's user. The system then deploys its multimodal interface, which "maps" the social space onto a representation of the relevant physical space, to locate the potential interlocutors and advise the user on an efficient communication strategy for the given community.

## 1   Motivation

Since the advent of computer age several decades ago, the role of various information systems in human knowledge sharing and proliferation has been increasing continuously. At the same time, however, the bulk of information learned by people in their lifetimes still never appears in a database or on the Internet but is readily available to members of various local communities, such as families, school students and alumni, indigenous people, company employee, and the like. This information is typically conveyed via word-of-mouth in conversations on an individual, person-to-person basis. While the modern information technologies traditionally focus on asynchronous mass-communication and deliver a vast array of tools (e.g. electronic libraries and search engines) supporting this form of information exchange, little has been done to assist the essentially personified and synchronous communication occurring daily, as we quire a teacher at a school, ask a local for directions, or seek advice from a friend or the "best" expert in a field (e.g. a doctor or lawyer). Even though existing computer systems do provide for person-to-person information exchange, their support does not go far beyond, say, a postal service that promotes communication among people who are already socially connected in one or another way. Whether we walk on a street or chat using an instant messenger, or else write to a forum of a social network system, our chances of obtaining information of interest are roughly the same. It is then our abilities to navigate social spaces (which are, at best, partly known) and to initiate and maintain communication at a level of synchronicity optimal for given time constraints that determine the success or otherwise of an information quest.

None of the present-day information systems and "e-services" known to the authors targets supporting this essentially "interhuman" navigation process. The very concepts of social space and communication synchronicity, although not totally alien in computer sciences, are presently discussed as quite theoretical and speculative rather than as something that would be practically used in and strongly affect information system design and development (Derene, 2008; Kalman & Rafaeli, 2007). While there is a growing interest to modeling social aspects of human communication and knowledge production processes in the relatively new fields of cognitive informatics and symbiotic computing, the community's present efforts are, however, mainly directed at the theory rather than at the development (see Wang & Kinsner, 2006).

Our study aims at the creation of an information service to facilitate human navigation in (unknown) social environments by enabling people to "perceive" and explore the corresponding social spaces. The envisaged service is also to help the users locate "carriers" of specific information (i.e. advisers) that would be approached in a particular situation. This paper describes a multi-agent information system "SoNa" (*So*cial *Na*vigator) developed to provide the social navigation service.

In line with the most common understanding of the social space concept (see Lefebvre, 1994; Monge & Contractor, 2003), the proposed system reproduces in a 3D virtual reality (a relevant fragment of) the physical space together with members of the local community present in the space at the moment. Unlike the physical proximity, social relationships (e.g. "trust" or "friendship") are usually not directly perceived in real life, but are inferred and "felt" from (collective and individual) communicative experiences. A haptic environment including a force display is then used to convey important parts of the community's communication practices – the "social knowledge" – to the user via the subconscious tactile communication channel. An agent network is created and used by the system to deal with the social knowledge. This network represents a real social network of the community, and the agents exchange information by communicating with their "socially connected" counterparts in the same way as people do it in the real world. Each agent in the network has parameters indicating whether the corresponding member is sociable, can be trusted, can afford to communicate (e.g. in terms of time), and is currently reachable (e.g. physically or via e-mail). Apart from exploration of the social space in various modalities and under different contexts, the user can use the system as a navigator in her or his search of a community member who would be approached with a specific information request.

In the next section, the design of the proposed system is presented. Section 3 describes a working prototype of the system implemented in the study. Section 4 elaborates on the developed multimodal user interface and user-system interactions. Section 5 then presents the haptic model of the social space. Section 6 gives an account of a case study of applying the proposed system's prototype in practice. Finally, Section 7 discusses related work and concludes the paper.

## 2 System Design

### 2.1 Overall Functionality and Architecture

Navigation in an environment, whether physical or virtual, can generally be defined as a four-stage iterative process (Spence, 1999): 1) perception of the environment, 2) reconciliation of the perception and cognition, 3) deciding on whether the goal has been reached, and 4) selecting the next action. Among these stages, only the first two directly depend on information about the environment and can thus be supported with an information system (Kryssanov *et al.*, 2002). To provide for navigation in a social space, the proposed system has five functions: 1) creating individual profiles of members of the community in focus and constructing a network of agents that reflects the community's communication and knowledge-sharing experiences, 2) displaying the social space as relevant fragments of the agent network together with established communication practices in 3D graphics and haptic virtual environments, 3) receiving the user's request and gathering the agents' knowledge, 4) extracting information that meets the user's needs, and 5) updating the states of the agents.

Figure 1 depicts the architecture of the system. The multimodal interface provides for the interaction of the user with other parts of the system and delivers information for the navigation process. The agent manager creates the agent network using data stored in the database. When the recommender system receives a request through the multimodal interface, it selects an appropriate agent and sends a query to this agent to gather relevant information in the network. For information gathering, an efficient agent-based communication algorithm proposed by F.E. Walter with co-authors is implemented (Walter *et al.*, 2008). Once the recommender system receives responses from all the agents in the network, it analyzes the obtained data and sends results of the analysis to the multimodal interface. The multimodal interface shows carriers of information sought by the users (who are thus the potential advisers) in the related segments of the social and physical spaces. The interface also assists the user in selecting among the advisers and in finding the "best approaching / communication strategy" for a given adviser.

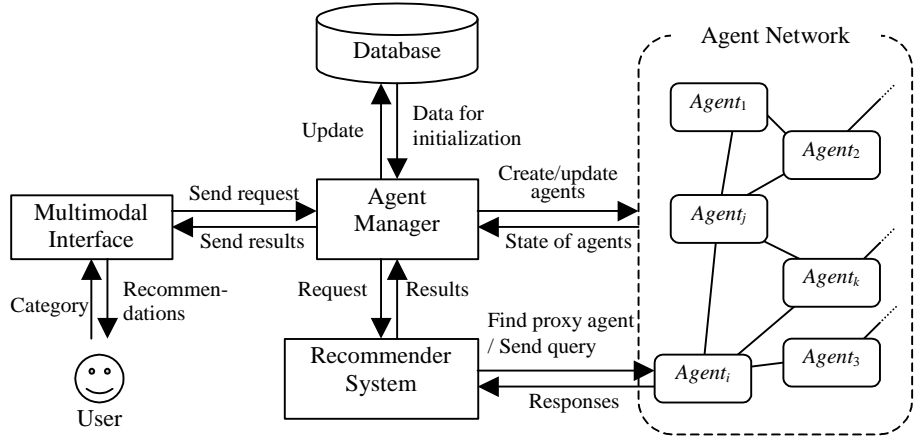

*Figure 1. System architecture*

## 2.2 Agent Network

The agent network is composed of the same number of agents as the number of members in the community in focus, and the agents are connected to their "acquaintance agents" just as the corresponding people are socially connected in the real world. To assemble the agents into a network, the following information is required: profiles with individual characteristics of the community members, personal network data of the community members, and each member's communicative experiences in respect to other members.

First, the agent manager creates $N_A$ agents, denoted $A = \{a_1, a_2, ..., a_{N_A}\}$ (it is understood that $N_A$ is equal to the number of members in the community). Each agent's profile, which is registered in the database, initially includes only static data, such as member's name, gender, and usual ("permanent") location/address in the physical space. This static data is set as the agent's parameters. Next, the agent manager informs agents (by attaching relevant descriptions – keywords) about the expertise of the corresponding members in the real world. Members with expertise are specified as $O = \{o_1, o_2, ..., o_{N_O}\}$, $N_O \leq N_A$. With $O$, we will thus denote members who have been evaluated by other members of the community in respect to a specific category of knowledge $c_k$. The evaluation is performed by rating the members as "relevant" or "irrelevant" when information sought falls into the category $c_k$. Totally, there are $N_C$ pre-defined knowledge categories denoted $C = \{c_1, c_2, ..., c_{N_C}\}$. The rate of $o_j$, $j \in [1, N_O]$, is represented as $r_j$. In the current implementation of the agent model, we assume the binary rate: $r_j = 1$ for the positive evaluation, and $r_j = -1$ otherwise.

Each agent is connected to its "acquaintance agents," using the members' personal network data. This data has the structure of an undirected graph, and the state of linkage between the nodes (i.e. agents) is represented, using an adjacency matrix. The matrix is constructed via mutual certification of pairs of agents. Each agent then receives two dynamic attributes: agents-acquaintances called "neighbors" and "trust" (that can also be thought of as "practical usefulness") values for the neighbors. A trust value between agents $a_i$ and $a_j$ is expressed as $T_{a_i,a_j}$, which is a real number fluctuating between 0 (no information / low trust) and 1 (full trust; if $i = j$, then generally $T_{a_i,a_j} = 1$). The dynamics of $T_{a_i,a_j}$ is specified with the following equations ($t = 0$ corresponds to the moment when the agent network is initialized, and each discrete time-instant $t > 0$ – to a consecutive update of the rating data in the member profiles; initialization by setting $T_{a_i,a_j}(t=0) = 0$):

$$\tilde{T}_{a_i,a_j}(t=0) \equiv 0 \ , \tag{1a}$$

$$\tilde{T}_{a_i,a_j}(t+1) = \begin{cases} \gamma \tilde{T}_{a_i,a_j}(t) + (1-\gamma) r_k, & r_k > 0 \\ (1-\gamma) \tilde{T}_{a_i,a_j}(t) + \gamma r_k, & r_k < 0, \end{cases} \tag{1b}$$

$$T_{a_i,a_j}(t+1) = \frac{1 + \tilde{T}_{a_i,a_j}(t+1)}{2}, \quad t = 1, 2, \ldots \ , \tag{1c}$$

where $r_k$ is the product of rates by $a_i$ and $a_j$ in respect to member $o_k$, and parameter $\gamma$ determines to what extent the previous trust value affects the new trust value. When $\gamma$ is greater than 0.5, the trust value increases slowly, and decreases quickly that is the "trust dynamics" usually observed in real social networks (Walter *et al.*, 2008).

The trust value between two agents that are not directly connected (i.e. are not neighbors) is calculated as the product of all of the trust values in the query path connecting the two agents. At any time, members can add new rates for other community members into the knowledge of their respective agents. The calculated trust values are used by the recommender system when there are more than one person to recommend under the same conditions. The data of a member recommended by $a_j$ is sent to the user interface with a weight $\omega$ calculated for the user's agent $a_i$ as follows:

$$\omega = \frac{e^{\beta \hat{T}_{a_i,a_j}}}{\sum_{k=1}^{N_R} e^{\beta \hat{T}_{a_i,a_k}}} \ , \tag{2}$$

where

$$\hat{T}_{a_i,a_j} = \frac{1}{2}\ln\left(\frac{1+2(T_{a_i,a_j}-0.5)}{1-2(T_{a_i,a_j}-0.5)}\right) . \qquad (3)$$

In formula (2), $N_R$ is the number of responses to the specific request made by $a_i$, and parameter $\beta$ determines to what degree the weight $\omega$ accounts for the trust value $T_{a_i,a_j}$: when $\beta$ is 0, all weights are the same, and when $\beta$ is close to 1, a member with a higher trust value receives a greater weight.

### 2.3  Agent Functions and the Recommender System

Whenever the system user is not a member of the community, the recommender system accesses the agent network to find an agent with a profile most similar to the user's self-description, and makes this agent the user's proxy. The user, whose agent (or proxy) is $a_i$, inputs a category of her/his enquiry, $c_k$. The recommender system sends the user's query to $a_i$ in the agent network, where it is relayed by $a_i$ to its neighbors as $query(a_i, c_k)$. When a neighbor of the agent receives the query, it checks if it has knowledge about members rated in category $c_k$. If the neighbor agent finds a rated member, it sends a response to the agent $a_i$. The response is formed as $response(a_i, a_j, c_k, (o_l, r_l), T_{a_i,a_j})$, where $a_j$ stands for the agent, which sent the response, and $o_l$ is the member, which is rated with $r_l$ by $a_j$ in category $c_k$; $T_{a_i,a_j}$ is the trust value between $a_i$ and $a_j$. If the neighbor does not have knowledge about relevant members for the given category of expertise, it further transmits the query to its neighbors. These latter neighbor-agents process the query in the same way as described above. To prevent unnecessary communications, every agent, which has once processed a query, ignores this query when it is received repeatedly. Usually, there are many paths between agents $a_i$ and $a_j$ in the agent network, but the generated responses always pass through the same path as the query does.

When the information-gathering algorithm terminates, the agent, which originally sent the query, has a set of responses from the community members. The next step is selecting members, which meet the user's criteria, to recommend from the available set. In the set, there may be people impossible to approach at the given moment, regardless of how strongly other community members would recommend them. For example, when the user needs to meet an adviser immediately, if the system recommends a person who is currently not reachable or who can only speak a foreign language not mastered by the user, such a recommendation would have little practical value. The system filters the responses to remove any potentially useless recommendations and, while doing so, attempts to balance the synchronicity of the expected communication.

## 3    Developed Prototype

In our study, we reconstructed the social network of the Intelligent Communication Laboratory, College of Information Science and Engineering, Ritsumeikan University. For a typical application of the system, we considered situations where a student having troubles with studying particular subjects, such as networking or programming, seeks an advice for her/his study. She/he thus needs to find out who would be the best candidate—a member of the laboratory—to be requested for help.

The modeled community is composed of 43 members. Each member created a profile containing the member's name, gender, grade, and certain dynamic characteristics, such as personal network data, availability, and trust values calculated via rating other students in the laboratory. The rating data has been collected from the laboratory members in regard to some 19 categories about classes (e.g. Math), specialized knowledge topics (e.g. Java programming), and the campus life (e.g. Events). The members were asked to select and rate maximum three other members, whom they previously requested for help in the given category. By implementing the information-gathering algorithm described in the previous section, the system's developed prototype can then choose the "best" adviser in the category, as it is socially recognized in the community.

## 4    Multimodal Interface and User-System Interactions

The proposed system has a multimodal interface to facilitate the system-user interaction and increase the efficiency of the navigation process. In SoNa, the user receives information about the community not only in a visual form, but also from physically sensing objects and areas in the virtualized social space. For the latter, the user manipulates a haptic device – force display PHANToM (see http://www.sensable.com/; last accessed on March 9, 2009). As the force display reproduces the reaction force, the user receives additional information about the friendliness and socializability "structure" of the community. In addition to representing the community's physical disposition in the "traditional" 3D virtual reality, the interface then delivers relevant social and personal information via the tactile channel by imposing force fields on the 3D graphics space. Figure 2 illustrates modalities of the user-system interactions for the developed prototype (in the figure, gray blocks indicate active channels).

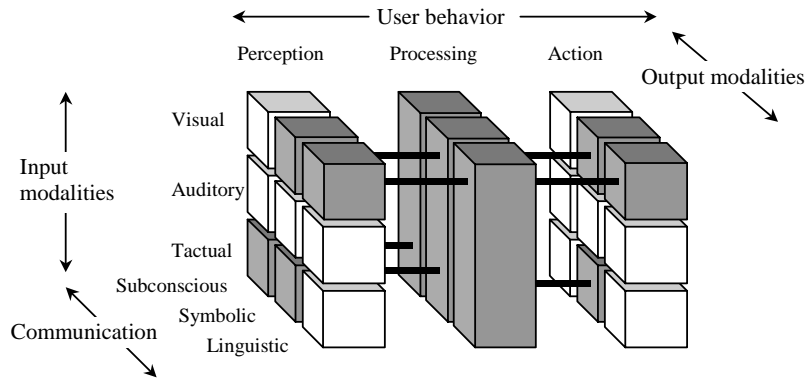

*Figure 2. Exploration of the social space with the multimodal interface*

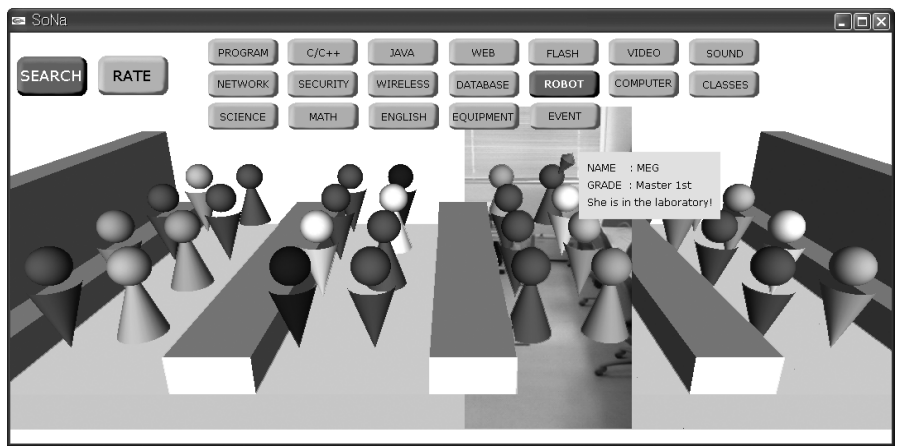

*Figure 3. Screen-shot of the prototype interface*

Figure 3 is a screen-shot of the prototype's interface, where the members of the community are represented with cone and sphere polygons. The members are placed in a virtual space partly reconstructing the laboratory settings with relevant photographic images interactively displayed for locations currently under exploration. In the screen-shot, the "active" area around the cursor is the two lines—the $3^{rd}$ and $4^{th}$ from the right—of students highlighted with the photographic image window in the background. The shape of the displayed member representation depends on the member's gender, and the color—on the grade. When the user "touches" a member in the virtual reality with the haptic interface pointer (HIP), the corresponding user profile appears. If the user touches the "Search" button on the screen, new 19 buttons pop up and are used to specify a category of the user's request. Once a category is selected, the agent of the user attempts to find the best advisers, utilizing the agent network. The potential advisers are proposed by the recommender system, and the best 3 chosen members are indicated with brighter color tones, while scalar force-fields are reproduced with the haptic device to assist the user's navigation in the social space created by the members currently present in the displayed physical space.

## 5  Haptic Environment

### 5.1  Social Space Model

The social space is composed of the community members, where each member is represented in the tactile space as a solid object. Tactile characteristics of the represented members are set to reflect their social qualities. An underlying assumption of the proposed model is that different social qualities can be associated with and evoke different emotions which, it turn, can be associated with tactile perceptions. The model assumes the following very general associations: "friendliness" $\rightarrow$ "pleasant emotions," "socializability" $\rightarrow$ "anxiety/elevated emotional states," and "trustability" $\rightarrow$ "pleasant or neutral emotions." To extend these associations to tactile representations, an experiment was conducted, in which subjects were asked to try to describe their communicative emotional experiences (emotions) in terms of the following tactile perceptions: "hard or soft," "sticky or slick," "rough or smooth," and "attracting or repulsing."

Totally, 224 subjects participated in the experiment, with 47% females and 53% males, all Japanese adults aged from 19 to 45 years old. The collected emotional experiences were classified into 5 general, non-orthogonal emotion classes, as it is shown in Table 1, using the emotional state model proposed in (Desmet, 2002). Table 2 shows results of the experiment.

No statistically significant difference was detected for the answers by males and females; the data also appeared statistically homogeneous in respect to the respondent's age. Among the statistically significant results ($p<0.05$), we would like to point to the following. The pleasant/unpleasant dimension is, apparently, well associated with an attracting/repulsing force filed. Pleasant emotions are also associated with softness, while unpleasant with stickiness (i.e. high viscosity). For the anxiety/boredom dimension, roughness (as opposite to smoothness) appears the dominant association.

Based on the experimental results, the following social space model was implemented in the system prototype. All the laboratory members (as well as other objects of the physical environment, e.g. furniture, etc.) are represented as solid, by default hard objects. The stiffness coefficient is adjusted in proportion to the member's friendliness so that members called by the other members as "friends" appear softer. At the same time, a viscosity field is generated around an object representing an "unfriendly" (i.e. declared by few or none as friend) member who is, nevertheless, chosen as adviser by the recommender system. The roughness of the represented object surface is set to a high value for members with high socializability. Attraction/repulsion haptic force fields are generated around objects representing members selected by SoNa as advisers in the given situation, where the driving force is adjusted in proportion to the trust value between the user's agent and the agent of the adviser closest (in the physical space) to the position of the HIP. With a threshold set for the trust parameter at 0.5, the lower trust values trigger generation of repulsion, and the higher values – of attraction fields.

*Table 1. Human emotions classified in regard to their "value" (pleasant/unpleasant) and psychological arousal (anxiety/boredom)*

| Class | Emotions |
|---|---|
| Unpleasant emotions | Disgusted, frightened, annoyed, hostile, despise, indignant, alarmed, irritated, frustrated, bewildered, nervous, contemptuous, disturbed, flabbergast, jealous, aversive, irked, moody, grouchy, ashamed, cynical, embarrassed, disappointed, dissatisfied, disapproving, confused, gloomy, melancholy, isolated, sad, guilty, disillusioned, bored |
| Pleasant emotions | Loving, jubilant, excited, desiring, inspired, enthusiastic, entertained, admiring, joyful, fascinated, yearning pleasantly, proud, surprised, happy, appreciated, amused, cheerful, sociable, attracted, fulfilled, intimate, satisfied, cozy, comfortable, softened, relayed |
| Anxiety | Disgusted, frightened, annoyed, hostile, despise, indignant, alarmed, irritated, frustrated, bewildered, nervous, loving, jubilant, excited, desiring, inspired, enthusiastic, surprised, concentrated, eager, astonished, amazed, aroused, longing, avaricious, curious |
| Boredom | Gloomy, melancholy, isolated, sad, guilty, disillusioned, bored, fulfilled, intimate, satisfied, cozy, comfortable, softened, relayed, composed, awaiting, deferent, passive |
| Neutral emotions | Surprised, concentrated, eager, astonished, amazed, aroused, longing, avaricious, curious, composed, awaiting, deferent, passive |

*Table 2. Human emotions and the associated tactile perceptions (for each perception, the number in the table indicates the number of subjects with the given association)*

| Emotion class | Hard | Soft | Sticky | Slick | Smooth | Rough | Attracting | Repulsing |
|---|---|---|---|---|---|---|---|---|
| Unpleasant | 89 | 0 | 162 | 53 | 33 | 191 | 10 | 153 |
| Pleasant | 4 | 193 | 11 | 103 | 101 | 2 | 152 | 22 |
| Anxiety | 64 | 20 | 73 | 17 | 9 | 97 | 112 | 62 |
| Boredom | 124 | 156 | 11 | 117 | 156 | 28 | 11 | 55 |
| Neutral | 104 | 4 | 24 | 41 | 63 | 4 | 35 | 22 |

**5.2   Haptic Force Modeling**

The described prototype was implemented in C++ programming language with its haptic environment built on top of the SmartCollision Studio™ commercial software package (Tamura *et al.*, 2007). The latter package provides penetration depth calculation in real-time with realistic friction and elastic force modeling during collisions of the HIP and (visualized) geometrical objects. The stiffness (or "solidness") coefficient was associated with personal "friendliness" of displayed members who are not among the 3 best advisers (e.g. "hard" stands for an "unfriendly" subject). After a short training, the participants of our case study (see Section 6) could easily discriminate "soft", "average," and "hard" virtual objects for the stiffness coefficient set at 75, 200, and 350 N/m, respectively. The friction (i.e. "roughness") coefficient was associated with the member's socializability (higher friction—better socializability—easier to perceive/harder to miss or overlook). Usually, the feedback force is set to zero when there is no collision of the HIP with objects. However, this method does not allow for discriminating between intuitive, distant perceptions of various members and of the empty (i.e. "member-free") space. For the exploration of the entire social space, we introduced a new approach based on scalar viscosity fields set around recommended but "unfriendly" members, and attracting/repulsing force fields – around "trustable/untrustable" recommended members, as it is stipulated by the social space model described in the previous section.

The dynamic model of the method is described as follows. Let us first consider a point of mass $m$ in a viscosity field $\lambda$. We will assume that the point is loaded by external "attraction" and driving forces, $F_a$ and $-F_h$, respectively. The point dynamics is then defined with the following differential equations:

$$m\frac{d^2\rho}{dt^2} + \lambda(\rho - \rho_1)\frac{d\rho}{dt} = F_a(\rho - \rho_0) - F_h(t) \ , \tag{4a}$$

$$F_h(t) = k_h \Delta\rho(t) + b_h \Delta\frac{d\rho(t)}{dt} \ , \tag{4b}$$

where $\rho = (x, y, z)^\mathrm{T}$, T is the transposition operator, is the point radius-vector. The driving force is opposite to the haptic feedback force $F_h$, which is calculated in real-time by the standard "spring-damper" model (Goncharenko *et al.*, 2006, 2007), using the PHANToM coordinate input. In equations (4), $\Delta\rho(t)$ is a vector from the current HIP position to the mass point, and $k_h$ and $b_h$ are coefficients of the spring-damper model. For simplicity, we used only one "attraction/repulsion" pole at position $\rho_0$, and calculated $F_a$ as the force in the direction to (or from, in the case of repulsion) $\rho_0$ and proportional to the distance between $\rho$ and $\rho_0$. Likewise, we selected only one focus of "unfriendliness" at $\rho_1$ and calculated an isotropic scalar viscosity field $\lambda$, which depends on the distance to the focus of "unfriendliness" as

$$\lambda(\rho - \rho_1) = \frac{c_a}{1 + d_a(\rho - \rho_1)^2}$$, where $c_a$ and $d_a$ are some constants.

In the beginning of haptic interaction, it is assumed that the HIP and the mass point coincide. The corresponding differential equation (4a) is numerically solved by a fourth-order Runge-Kutta method, using the real-time control function $F_h(t)$. In our study, the model parameters $c_a$ and $d_a$ were set to provide the viscosity from 1 kg/s to 15 kg/s inside the working PHANToM space.

The physical values of the haptic model coefficients were set as linearly proportional to the social characteristics obtained from profiles of the members. The proportionality constants were adjusted to yield subjectively smooth haptic feedback, which obviously makes it difficult to move around an "unfriendliness" pole, and which "guides" in the direction to the focus of attraction but (weakly) drives from the focus of repulsion. The personal friendliness was set proportional to the total number of people declaring the member as "friend," and the socializability – to the number of contacts in the member's personal network. In all cases, the pole location is defined by the nearest (in respect to the HIP) recommended member's position in the physical space. The dynamic model parameters were adjusted to yield negligible forces and viscosity when the physical distance to the nearest recommended member exceeds the "social distance" of 2 meters (Hall, 1966). Model (4) with one attracting (or repulsing) pole and a central radial viscosity field provides for subjectively good intuitive reinforcement guidance in the social space. It was found experimentally that haptic guidance becomes ambiguous when the number of poles is more than two. Therefore, only the recommended members nearest to the HIP are visualized in the tactile space with the force fields (the latter, however, results in some "tactile discretization" effects when perceiving the social space with the haptic device).

## 6  Case Study

The developed prototype was installed on an ordinary desktop computer (with the haptic device connected) in the Intelligent Communication Laboratory. The laboratory members were asked to review and possibly update their profiles in the system database at least once a week during the spring semester when the case study took place. The laboratory is also equipped with a semi-automatic system monitoring the current location of each member – this data was used to automatically update the location dynamic parameter of the agents in the prototype's agent network.

*Table 3. Advisers selected by group A*

| Category | 1st adviser | 2nd adviser | 3rd adviser |
|---|---|---|---|
| C1 | Aaron (6) | Bill (6) | Carl (3) |
| C2 | Eddy (9) | Aaron (3) | Bill (3) |
| C3 | George (12) | Dave (3) | -- |

*Table 4. Advisers selected by group B*

|  | Before using the system | | | After using the system | | |
|---|---|---|---|---|---|---|
| Category | 1st | 2$^{nd}$ | 3rd | 1st | 2nd | 3rd |
| C1 | Bill (9) | Carl (3) | Frank (3) | Bill (9) | Carl (6) | -- |
| C2 | Eddy (12) | Aaron (3) | -- | Aaron (12) | Eddy (3) | -- |
| C3 | George (6) | Aaron (6) | Dave (3) | George (15) | -- | -- |

*Table 5. Ranking obtained using questionnaires filled in by all the laboratory members*

| Category | The best adviser | 2$^{nd}$ adviser | 3$^{rd}$ adviser | 4$^{th}$ adviser |
|---|---|---|---|---|
| C1 | Carl | Bill | Aaron | Frank |
| C2 | Aaron | Eddy | Henry | Ian |
| C3 | George | Dave | Aaron | Jack |

Thirty students who are not members of the laboratory and are generally unfamiliar with the community were selected as subjects, which were then divided into two groups: A and B. The subjects were asked to name a member of the laboratory who, in their opinion, would be the best adviser in three knowledge categories. The categories were randomly assigned from the 19 categories supported by the developed prototype. The subjects in group A were given 10 minutes to decide on the best adviser in each category, based solely on their own communicative experience. The subjects in group B were asked to first use only their own experience and then the system, and name the best advisers in both cases. To reduce the possible bias, the B-group subjects were told that the system does not always recommend the best adviser, and that they should rather rely on own communicative experience in choosing a candidate for the contact. The total (communication, user-system interaction, and decision-making) time for every subject in group B was limited by 6 minutes. Prior to the experiment, group B was familiarized with the user interface of the prototype but received no explanations about the specific "meaning" of the reaction force feedback in the haptic environment. Subjects in both groups were encouraged to communicate any member of the laboratory at any moment within the given time frames. Tables 3 and 4 show results of the evaluation of the members (with private information removed), where the number in the parentheses is the number of subjects who contemplated the given member as a useful adviser.

Based on the presented results, two points can be discussed: the effectiveness and the efficiency of SoNa. After using the system, all the subjects belonging to group B changed their answers in one or two categories. All the changes made were related to higher-ranked members. Although all the subjects in both groups were given almost the same time (10 and 6 minutes) to think about the questions, the members who used the system chose statistically "better" ($p<0.05$) advisers, as it can be inferred by comparing the individual answers with the entire community's knowledge shown in Table 5. It can, therefore, be said that the system is effective and efficient, as it indeed provides the user with the best information available and, at the same time, does not increase the decision-making time.

# 7 Conclusions

The experimental results obtained in the case study suggest that the proposed system can be a useful tool for assisting human navigation in unknown social environments, and for increasing the efficiency of information search in such environments. It is understood, however, that while the developed system is a working prototype that can be used in practice "as is," at least some of the design solutions implemented in the multimodal interface are rather arbitrary. Larger scale and more elaborated experiments need to be conducted to justify the choice of specific parameters of the social network, which are used to reconstruct the social space in the virtual reality. Although the presented system has, to the authors' best knowledge, no direct analogs, some of the recently reported design solutions would be used to enhance the proposed service concept and facilitate the social space reconstruction. Specifically, to improve the currently deployed profile-based model of the community's social network, a human activity tracking mechanism, as it was proposed in other work (Wojek *et al*., 2006; Danninger *et al*., 2005), would be used to assess communication practices of the community in a closed (laboratory-like) environment. Besides, the development of the community's knowledge model, e.g. an ontology as was proposed in (Giménez-Lugo *et al*., 2005), would allow us to improve the recommender function of the system and extend its potential applicability to the domain of organizational management.

A deficiency in the authors' current understanding of the developed system utility is the role of the subconscious tactile interaction channel. As an audio interaction channel would naturally be added to the multimodal interface (that is, in fact, part of the authors' plans for future work), separate experiments should be conducted to analyze how the interactions in different modalities affect the navigation process. Although there have been earlier attempts to convey human emotions via the tactile channel (Picard, 1995; Smith & MacLean, 2007), the immediate communicative effect of the reproduced tactile perceptions as well as the place of "tactile" emotions in the social space remains an open research question (Haans & Ijsselsteijn, 2006).

The main contributions of the presented work, as seen by the authors, are the original concept of the social navigation support service, the haptic model of the social space, and the developed information system that realizes the envisaged theoretical concepts. Specific design solutions may and will be changed, however. In the next version of the prototype, we plan to significantly increase the size of the agent network and the number of knowledge categories supported. Attempts will also be made to improve the scenes reproduced in the 3D graphics virtual reality along with the haptic image of the social space.

## Acknowledgments

The authors would like to thank all the members of the Intelligent Communication Laboratory for their participation in the case study. The contribution of Izumi Kurose, who collected the emotional association data for the social space model, is gratefully acknowledged. Discussions of the social space model with Eric W. Cooper are also appreciated.

## Authors' Biographies

Victor V. Kryssanov received his MS in physics from the Far-Eastern National University, Russia, and PhD in biophysics from the Russian Academy of Sciences in 1991 and 1994, respectively. He currently serves as Professor of Information and Communication Science at the College of Information Science and Engineering, Ritsumeikan University, Japan.

Shizuka Kumokawa received her BS in engineering from Ritsumeikan University in 2008, and she is currently pursuing an MS degree in information science at Graduate School of Science and Engineering, Ritsumeikan University.

Igor Goncharenko holds an MS in control systems from the Moscow Institute of Physics and Technology (1984) and a PhD in computer science from the Russian Academy of Sciences (1994). He is currently a Chief Research Officer of 3D Incorporated, Yokohama, Japan.

Hitoshi Ogawa received his Dr.Eng. from Osaka University in 1977. He is currently Professor at the Department of Information and Communication Science, College of Information Science and Engineering, Ritsumeikan University.